# Spin-selective tunneling from nanowires of the candidate topological Kondo insulator SmB$_6$

Anuva Aishwarya[1], Zhuozhen Cai[1], Arjun Raghavan[1], Marisa Romanelli[1], Xiaoyu Wang[2], Xu Li[3], G. D. Gu[4], Mark Hirsbrunner[1], Taylor Hughes[1], Fei Liu[3*], Lin Jiao[1,2*] and Vidya Madhavan[1*]

[1]Department of Physics and Materials Research Laboratory, University of Illinois Urbana-Champaign, Urbana, IL 61801, USA

[2]National High Magnetic Field Laboratory, Florida State University, Tallahassee, FL 32310, USA

[3]State Key Laboratory of Optoelectronic Materials and Technologies, Guangdong Province Key Laboratory of Display Material and Technology, and School of Electronics and Information Technology, Sun Yat-sen University, Guangzhou 510275, China

[4]Condensed Matter Physics and Materials Science Department, Brookhaven National Laboratory, Upton, NY 11973, USA

## Abstract

**Incorporating relativistic physics into quantum tunneling can lead to exotic behavior such as perfect transmission via Klein tunneling. Here, we probe the tunneling properties of spin-momentum locked relativistic fermions by designing and implementing a tunneling geometry that utilizes nanowires of the topological Kondo insulator candidate, SmB$_6$. The nanowires are attached to the end of scanning tunneling microscope tips, and used to image the bicollinear stripe spin-order in the antiferromagnet Fe$_{1.03}$Te with a Neel temperature of ~50 K. The antiferromagnetic stripes become invisible above 10 K concomitant with the suppression of the topological surface states. We further demonstrate that the direction of spin-polarization is tied to the tunneling direction. Our technique establishes SmB$_6$ nanowires as ideal conduits for spin-polarized currents.**

A direct consequence of the non-trivial topology of $Z_2$ topological insulators (TIs) is the existence of an odd number of spin-momentum locked boundary modes with Dirac dispersion *(1)*. When the Fermi energy is tuned inside the bulk band gap, transport is dominated by the Dirac surface states and striking topological phenomena can be observed. In this scenario, spin-momentum locking coupled with Dirac dispersion dictates that the net current traveling in a given direction must be spin-polarized; currents traveling in opposite directions carry opposite spin. The high spin-charge conversion efficiency, and the topological protection of the boundary states, make TIs particularly interesting for applications in "topological spintronics" as well as for quantum technologies such as controlling qubits *(2-4)*. To date, many exotic phenomena have been realized in magnetically doped TIs or TI heterostructures, such as the quantum anomalous Hall effect *(5)*, topological magnetoelectric effect *(6)*, and large spin–orbit torque effect *(7)*. TI boundary modes are also ideal platforms to explore the unique tunneling properties of relativistic fermions.

In practice, a substantial bottleneck to using TIs both for fundamental studies and applications is the fact that many semiconductor-based TIs suffer from intrinsic defects that cause self-doping, which moves the Fermi energy into the bulk-bands. There has therefore been a tremendous impetus to study interaction-driven topological phases where the Fermi level is naturally pinned to the topological band gap. To this end, topological Kondo Insulators (TKIs) have been proposed as ideal topological systems, where surface Dirac fermions naturally dominate the density of states near Fermi level *(8)*. In TKIs, the lattice Kondo effect leads to the emergence of a low-energy, correlation-driven heavy band, which hybridizes with another light band crossing the Fermi energy, leading to a Kondo hybridization gap *(9)*. A concomitant band inversion results in the formation of topological surface states (TSS) within this gap. The material $SmB_6$ has been proposed as a TKI and exhibits the following phenomenology *(9)*. At high temperatures, $SmB_6$ is a Kondo metal. Below the bulk Kondo coherence temperature ($T_K$) of ~50 K, the hybridization between the emergent heavy f-band and the light d-bands opens a Kondo gap (schematic of Fig. 1B), which becomes better defined with decreasing temperature *(10-14)*. Scanning tunneling microscopy (STM) studies have shown that below 10-15 K, ("heavy") surface states develop within the Kondo gap *(15,16)*. The observation of the surface states at far lower temperatures (~10 K) than the bulk $T_K$ suggests that the surface Kondo temperature may be lower than the bulk *(15,17)*. The topological nature of the surface states is supported by the observation of a series of phenomena *(18)*, e.g., non-local transport *(19,20)*, enhanced spin-pumping *(21)*, the Edelstein effect *(22-23)*, chiral edge modes *(24)*, counter-clockwise spin textures for the surface electron band *(25),* and the Klein paradox *(26)*. Despite these studies, the topological origin of the surface states continues to remain controversial *(27)*. If the surface states were indeed topological, the presence of isolated TSS within a bulk band gap that straddles the Fermi energy, would make $SmB_6$ an excellent system for harnessing the properties of Dirac surface states.

In this work, we use a unique tunneling geometry to reveal a distinctive tunneling process that we dub helical tunneling. The term `helical' refers to a direction-dependent spin polarization akin to the spin-momentum locking of helical edge states of a quantum spin Hall insulator. We show that the tunneling electrons from $SmB_6$ nanowires are spin-polarized, with the direction of spin-polarization being tied to the direction of tunneling. To achieve this, we use focused ion beam (FIB)-based nanofabrication techniques, to harvest and attach $SmB_6$ nanowires (NWs) to the ends of etched tungsten wires typically used as STM tips (see Fig. S1) *(28)*. The $SmB_6$ nanowires used in this work were grown on silicon (Si) substrates. They have diameters ranging from 60 nm to 100 nm, and average lengths of ~50 microns (Fig. 1C and inset, and 1D). The procedure for the growth of the nanowires is described in detail elsewhere *(29)*. The NWs grow along the [001] direction and terminate laterally at high symmetry facets such as the (001). The larger diameter

NWs we use typically contain multiple large facets where the electronic structure is expected to be similar to the surfaces of bulk $SmB_6$. Crucially, given that the penetration depth of TSS in $SmB_6$ is less than 30 nm, our $SmB_6$-NWs are sufficiently thick to host TSS *(21,26)*.

Mounting $SmB_6$-NWs on tungsten tips is a two-step process that uses the omni probe inside the FIB + SEM dual beam system *(28)*. The NW-tips thus prepared are transferred to the STM (Fig. 1E shows a schematic of the tunneling setup). For benchmarking, we first characterize the NW tips and their topological surface states on a Cu (111) single crystal sample (see Fig. S2 and S3) *(28)*. Using the fact that the STM d*I*/d*V* spectrum is a convolution of tip and sample density of states and the fact that Cu(111) has a flat density of states near the Fermi energy, we can isolate the density of states of the NW-tip. The data verify that the NW-tip hosts surface states, similar to bulk $SmB_6$. This indicates that the NWs inherit the surface and bulk properties of bulk $SmB_6$ crystals.

Based on these observations, we expect the surface states to play a key role in the tunneling characteristics of the NW-tips. In particular, the spin-momentum locking of topological surface states would result in spin polarized currents. To determine the spin-polarization of the tunnel current, we study iron telluride ($Fe_{1+x}Te$), which is a bicollinear antiferromagnet (AFM) for $x<0.11$ (Fig. 2A). $Fe_{1+x}Te$ has been extensively studied by spin-polarized STM, and it is well established that the atomic scale AFM order can be observed with a traditional magnetic tip, where it presents itself as stripes with double the Te periodicity *(30-32)* (Fig. 2, B and C). Owing to matrix element effects, tunneling is favored between spin parallel states, and suppressed between spin anti-parallel states *(33)*.

For our STM studies, $Fe_{1+x}Te$ (nominal x=0.03) single crystals were cleaved at ~90 K before inserting into the STM head. To characterize the NW tip we obtain $dI/dV$ spectra. Shown in Fig. 2H is an average $dI/dV$ spectrum obtained on $Fe_{1+x}Te$ at 1.7 K. With $SmB_6$ on the tip we observe a peak that is on the opposite side of $E_F$ compared to spectra on bulk $SmB_6$ (see comparison with bulk spectrum in Fig. S4) *(28)*. This demonstrates the existence of the surface states on the tip. We now use this tip to obtain topographic images of $Fe_{1+x}Te$. In contrast to topography using a tungsten (W) tip (Fig. 2, and F), which shows the expected square Te-lattice of the top layer with a lattice constant of 3.8 Å, the images with the $SmB_6$-NW tips show stripes (Fig. 2E). The double periodicity of the observed pattern is the same as the spin modulation of the striped AFM order (Fig. 2G), and the image is consistent with spin dependent contrast seen in several previous STM studies using magnetic tips *(30-32)*. Note that this phenomenology was confirmed using multiple NW-tips (Fig. S5) *(28)*. Because it is well established that spin-dependent spatial contrast in STM images arises from the tunneling of spin polarized electrons, the observed spin contrast directly indicates that the $SmB_6$-NW tips act as a source of spin-polarized currents.

As a first check of the possible involvement of the surface states in generating the spin-polarized currents, we measure the temperature dependence of the observed spin-contrast. Previous transport and spectroscopic experiments on bulk $SmB_6$ have revealed the attenuation of the Dirac surface states in $SmB_6$ with increasing temperature, and a transition from surface-dominated to bulk-dominated regime around 10 K-14 K *(15,16,34,35)*. Figures 3, A to D, show the temperature dependent topographies of $Fe_{1+x}Te$ in the same field of view. The AFM order can be seen clearly at lower temperatures. However, at ~10 K the striped signature of the AFM order disappears, leaving behind only the atomic corrugation of the Te lattice. This is also clearly seen in the fast Fourier transform (FFT) (Fig. 3, E and F). We have reproduced similar results with two other NW-tips (Fig. S6) *(28)*. Critically, upon cooling down, the AFM order reappears (Fig. S7) *(28)*. Given that the Neel temperature ($T_N$) for $Fe_{1.03}$Te is around 50-60 K, the disappearance of magnetic contrast around ~10 K must be caused by a change in the NW tip. In fact, previous STM images obtained with magnetic tips continue to show AFM order all the way to 45 K *(30)*. In addition, we have investigated the same samples with canonical spin-polarized Cr tips, which continue to show clear AFM stripes at 10 K (see Fig. S8) *(28)*.

Although the temperature dependence suggests a causal relationship between the presence of the surface states and the spin-polarized contrast (Fig. S9) *(28)*, there may be other explanations for the temperature dependence. For example, metallic samarium is AFM in nature and has a $T_N$ ~ 14 K *(36)*, so a Sm cluster on the tip might create spin-polarized currents that mimic this temperature dependence. To distinguish between these scenarios, we explore how spin-momentum locked tunneling of Dirac electrons is different from typical magnetic tips. We calculate the spin-resolved transport coefficients for tunneling between a normal 2D metal and a 2D band of spin-momentum locked, singly degenerate linearly dispersing states (see schematic in Fig.4D). A Zeeman term was used to open a gap in the inner band of 2D Rashba states, which mimics tunneling from a 2D Dirac band. The transmission coefficient with an infinitesimal bias voltage is equivalent to the differential conductance, and as is usual in tunneling, the sign of the bias voltage dictates the direction of travel of the tunnel current. The net spin polarization of the tunneling electrons going from left to right ($I_{L-R}$) and right to left ($I_{R-L}$) is plotted in Fig. S10B *(28)*. The calculations show two important features: first, the tunnel current is indeed spin polarized and second, the spin-polarization reverses when the direction of tunneling is reversed (helical tunneling). Such a reversal of spin polarization is neither expected nor observed for conventional magnetic tips *(37),* suggesting an unambiguous way to test the involvement of the surface states in generating spin-polarized currents by checking if the currents are oppositely polarized at opposite bias.

From the discussion above, a smoking gun proof of the proposed helical tunneling would be the observation of contrast reversal of the observed AFM stripes at opposite tip bias voltages. Shown in Figs. 4, A and B, (also see Fig. S11) *(28)* are images in the same area obtained at positive and negative biases with a NW-tip. The linecuts of the two images (Fig. 4C) show a reversal of the AFM contrast, indicating that the opposite spin channel is selected by the tunneling electrons upon switching the sign of the bias voltage. To better establish that this contrast-reversal is pervasive over the whole image, we perform phase-referenced FFT *(38)* (See Fig. S12 for the analysis method) *(28)*. As shown in Figs. 4, F and G, we observe a negative value for the relative phase amplitude at $q_{AFM}$ establishing a $\pi$-phase shift of the stripes at opposite biases. This contrast reversal is robust against thermal cycling (Fig. S13) *(28)*. Moreover, as expected, such contrast reversal is not observed for spin-polarized Cr-tips (Fig. S14 *(28)* and Ref. *38*).

Combined with the temperature dependence, the contrast reversal at opposite bias voltages suggests that the topological surface states participate in generating the spin polarized tunneling currents observed in our experiment. We note that although we have different facets on the nanowire, we expect the junction to be formed at the bottom of one of the side crystal planes (see schematic in Fig. S15) *(28)*. This is important because an equal average over all facets would lead to zero net spin-polarization. Indeed, the macroscopic W-wire, which hosts the NW, is slid into the tip-holder and tightened by a set-screw which always causes a tilt, and this naturally results in the tunnel junction being formed at one of the side facets (see schematic in Fig. S15) *(28)*.

Further support for the involvement of the surface states is obtained by directly measuring how the low energy features associated with the $SmB_6$ surface states change across the AFM stripes. Shown in Fig. 2I is a spectroscopic linecut across one period of the AFM stripe. We find that the amplitude of the low energy features associated with the surface state changes periodically as the tip is moved across an AFM stripe. This change in intensity is a result of changes in the tunneling probability caused by the different relative orientations of the tip surface state electrons spin and the sample spin. We note that this scenario of TSS mediated helical tunneling will survive even with weak surface magnetism that has been observed in $SmB_6$ at mK temperatures *(24)*. It has been suggested that the TSS may mediate a ferromagnetic RKKY interaction among magnetic impurities on the surface giving rise to this surface magnetism *(39)*. However, because the Dirac surface states are generated by the non-trivial topology of the bulk, weak surface magnetism would simply open a small (< 1meV) gap at the Dirac point, leaving the remainder of the helical tunneling physics outside this energy window unchanged. This would not be at odds with any of our observations.

In summary, the reduced dimensionality and geometry of the NW-tips have allowed us to study the tunneling properties of a candidate TKI. Our observations and calculations show that the spin-polarization of the tunnel current is insensitive to the position of the Fermi energy relative to the Dirac point, and only depends on the direction of tunneling. These unique properties of spin-momentum locked Dirac fermions may directly be utilized for manipulating spin currents in the field of spintronics. In addition, magnetic contrast at surfaces or local defects can be detected with atomic resolution without using external magnetic fields, as would be necessary in a traditional spin polarized STM. Equally importantly, our technique of mounting nanowires on the STM tip opens up avenues for using topological NW tips for devices and heterostructures *(40,41)*. As an example, a proximitized topological nanowire in a vertical magnetic field would host Majorana fermions at the end effectively making a Majorana tip. This may then be used for directly studying interactions as a function of distance between two Majorana modes (one on the NW-tip and the other on the sample), which has proven difficult to do so far. Finally, scanning probe microscopy using unconventional tips is a burgeoning field in the study of quantum materials for nanoscale magnetic imaging *(42, 43)*. Our work thus opens many avenues for using various kinds of functionalized nanowires as STM tip (superconducting, topological, etc.) for atomic resolution imaging, in order to provide nanoscale information on different materials properties like magnetism, superconductivity and other emergent excitations.

**Acknowledgements:**

**Funding:** The STM work at the University of Illinois was supported by the National Science foundation through DMR-2003784. M.H and T.L.H thank ARO under the MURI Grant W911NF2020166 for support. V.M acknowledges partial support from the Gordon and Betty Moore foundation EPiQS grant #9465, and the Materials Research Laboratory Central Research Facilities at the University of Illinois for use of the cleanroom. V.M is a CIFAR Fellow in the Quantum Materials Program and acknowledges CIFAR for support. A.A. was partially funded by the PPG-Materials Research Laboratory Graduate assistantship 2020-2021. X.L and F.L. are very thankful for the support of the National Science Foundation of China (Grant Nos. 51872337), National Project for the Development of Key Scientific Apparatus of China (2013YQ12034506). L.J. and X.W. acknowledges financial support from National MagLab, which is funded by the National Science Foundation (DMR-1644779) and the state of Florida. The authors would like to thank Honghui Zhou, Senior Research Scientist, Materials Research Laboratory, University of Illinois for her help in using the FIB + SEM Dual beam system and acknowledge helpful discussions with Steffen Wirth and Peng Xiong.

**Author contributions:** A.A., L.J. and V.M. conceived the experiments. A.A., A.R. and L.J. obtained the STM data and characterized the tips and samples. L. J., M.R., and Z.C. fabricated the nanowire tips. G.D.G. provided the single crystals of $Fe_{1.03}Te$ and X.L. and F.L. grew the nanowires of $SmB_6$. X.W., M. H. and T.H. carried out the theoretical modelling. A.A., L.J. and V.M. performed the analysis and wrote the manuscript with help from all the authors.

**Competing interests:** The authors declare no competing interests.

**Data and materials availability:** All data have been uploaded to the Illinois Data Bank *(44)*.

# Figures

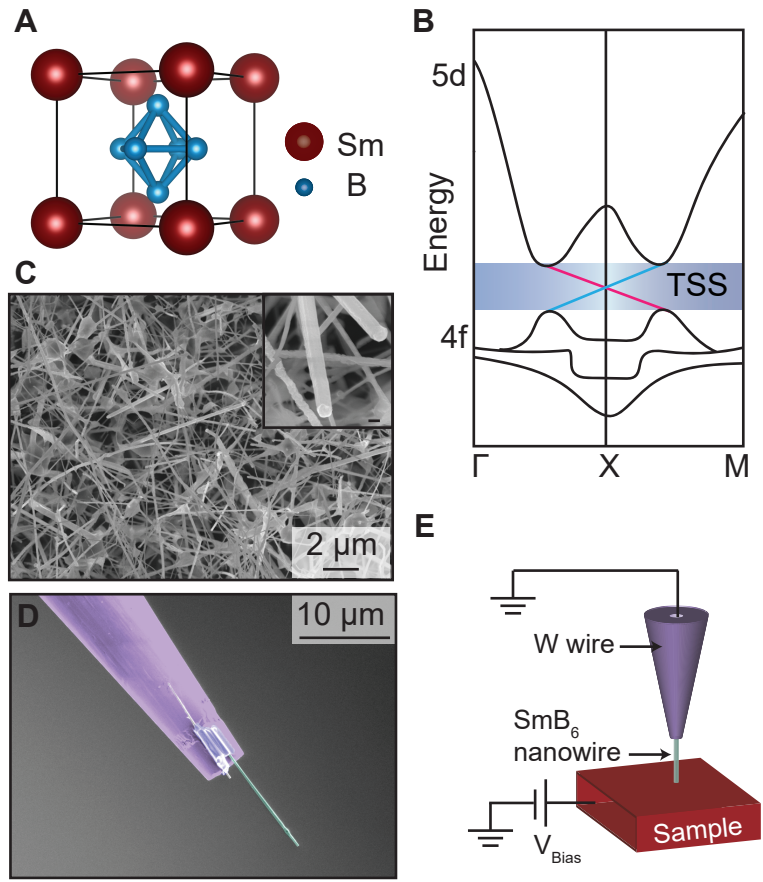

**Figure 1: Morphology, crystal structure, and electronic structure of the SmB$_6$ NW-tip**

**A,** Crystal structure of SmB$_6$ showing cubic lattice of samarium atoms with octahedrons of boron at the center. **B,** Schematic of the band structure of SmB$_6$ showing the hybridization induced gap between the 5d and 4f bands. The blue shaded region represents the Kondo gap, and the cyan and pink lines indicate the TSS. The colors represent opposite spin orientations. **C**, A high resolution SEM image of the nanowires on a silicon substrate. Inset: zoomed in image of a nanowire. Scale bar is 200 nm. **D**, A high resolution false colored SEM image of the nanowire tip showing the nanowire bonded to a trimmed tungsten tip. **E,** Schematic of the STM circuit showing the nanowire tip and the sample.

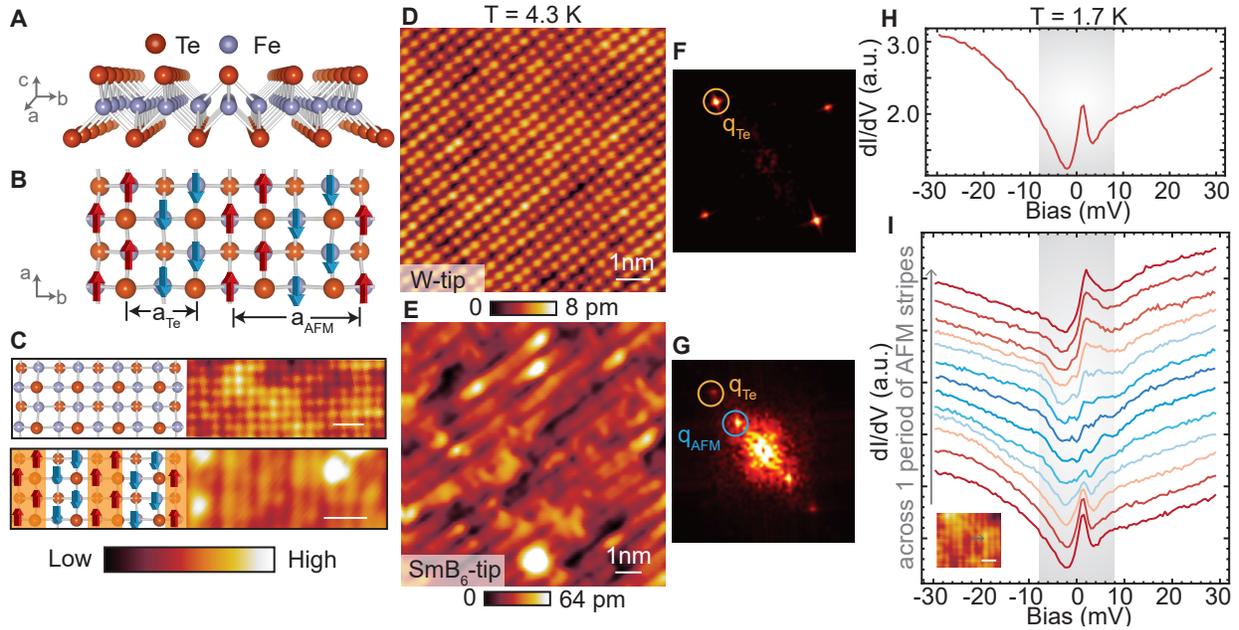

**Figure 2: Atomic resolution imaging of antiferromagnetic stripes in $Fe_{1+x}Te$ with $SmB_6$ NW-tip**

**A,** Crystal structure showing one layer of $Fe_{1+x}Te$ made of an iron (Fe) layer sandwiched between two tellurium (Te) layers. **B,** Schematic of the $ab$ plane of the lattice with the overlay of the spin structure showing the in-plane antiferromagnetic order. The periodicity of the antiferromagnetic order and the lattice are labelled as $a_{AFM}$ and $a_{Te}$, respectively. **C,** Schematic of the Te-lattice depicted by the orange balls (top) and AFM order (bottom). Topography as seen by a W-tip (top) and spin-polarized tip (bottom). The schematic and topography are not the same size. The scale bar in each topography is 1 nm. **D,** STM topography obtained with a spin-averaging tungsten (W) tip. ($V_{Bias}$ = 20 mV I = 60 pA, T = 4.3 K). **E,** STM topography obtained with the $SmB_6$ tip at $V_{Bias}$ = 20 mV, I = 60 pA, T = 4.3 K showing well defined spin-contrast. **F-G,** Fast Fourier transform of the topographies in **D,E** respectively showing only the Bragg peaks from the Te lattice ($q_{Te}$) (in **F**) and the presence of a strong signal from the antiferromagnetic ordering ($q_{AFM}$) along with the $q_{Te}$ (in **G**). **H,** Average dI/dV spectroscopy obtained on $Fe_{1+x}Te$ at T = 1.7 K. ($V_{Bias}$ = 50 mV I = 120 pA) In **H-I**, the highlighted grey region shows the feature in the DOS associated with the topological surface state. **I**, dI/dV linecut across one period of the AFM stripe (↑(red) to ↓(blue) to ↑(red)) showing the modulation of the TSS peak as a function of position. Inset shows the topography and the grey arrow shows the line where the spectra were obtained. Scale bar is 1nm.

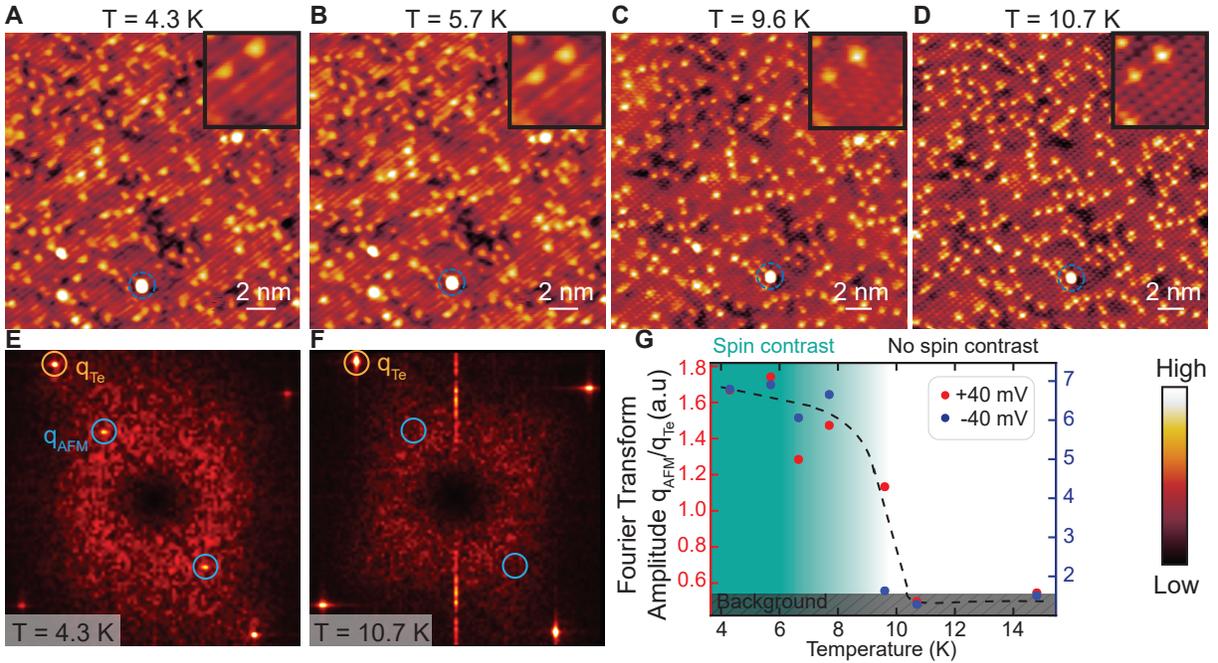

**Figure 3: Temperature dependence showing diminishing spin contrast**

**A-D,** STM topography on $Fe_{1+x}Te$ at the labelled temperatures, obtained in the same field of view, ($V_{Bias}$ = 40 mV, I = 40 pA). Blue circles denote the same defect in all frames. **E,F,** Comparison of FFTs of **A** and **D** to illustrate the disappearance of the $q_{AFM}$ at 10.7 K far below the antiferromagnetic $T_C$. A gaussian background has been subtracted from the center to improve visualization of the q-vectors. **G,** Plot of the ratio of the intensity of the signal from the antiferromagnetic ordering vector ($q_{AFM}$) to the Bragg peak ($q_{Te}$) as a function of temperature. Black dashed line is a guide to the eye. This shows that signal from the spin-contrast is highly suppressed at ~10 K, similar to the behavior of the topological surface state.

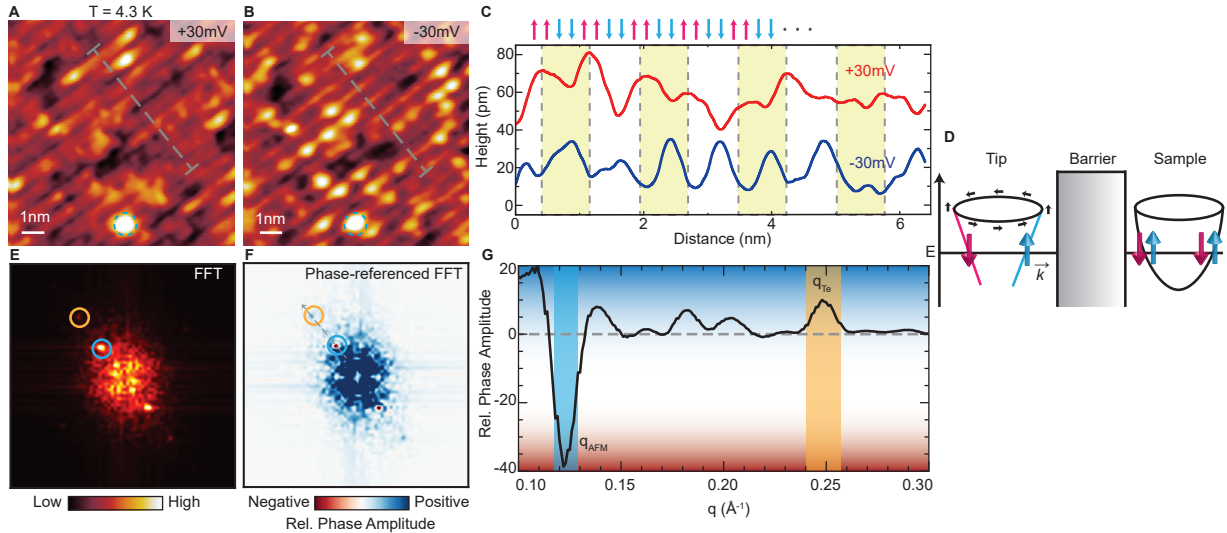

**Figure 4: Helical tunneling with SmB$_6$ NW tip: contrast reversal at opposite bias voltages**

**A** (**B**), STM topographies of Fe$_{1+x}$Te obtained in the same area at +30 mV (-30 mV) and using an SmB$_6$ tip at T = 4.3 K. **C**, Height profile of the topography along grey dashed lines in **A** (red curve) and **B** (blue curve). The red curve is shifted vertically for clarity. The dotted lines correspond to the peaks of the red curve and dips of the blue curve, indicating a $\pi$-phase shift of the stripes between the positive and negative bias scan. **D**, Illustration of the tunneling geometry used for the quantum tunneling simulation from the TSS to a normal metal. Without loss of generality, we have assumed the tip to be a 2D metal with just one branch of spin-momentum locked states crossing the Fermi energy and the sample to be a normal metal with spin-degenerate bands *(28)*. In reality, the sample has AFM stripes arising from the Fe lattice which is visualized by the spin-polarized currents from the tip. **E**, Absolute magnitude of the FFT of the topography shown in **B** showing the peaks associated with q$_{Te}$ (yellow) and q$_{AFM}$ (cyan). **F,** Phase-referenced FFT between the topographies shown in **A-B** with the same peaks highlighted in cyan and yellow. The relative phase amplitude is negative at $q_{AFM}$ indicating a $\pi$-phase shift of the stripes originating from helical tunneling. **G,** A linecut across the phase-referenced FFT (indicated by grey dashed line in F) illustrating the sign change between $q_{Te}$ and $q_{AFM}$.